\documentclass[preprint,showkeys,showpacs,superscriptaddress,nofootinbib]{revtex4}
  \usepackage{graphicx}
  \usepackage{rotating}
  \usepackage{multirow}
  \begin{document}
  \title{$J/{\psi}$ ${\to}$ $DP$, $DV$ decays in the QCD factorization approach}
  \author{Junfeng Sun}
  \affiliation{Institute of Particle and Nuclear Physics,
              Henan Normal University, Xinxiang 453007, China}
  \author{Lili Chen}
  \affiliation{Institute of Particle and Nuclear Physics,
              Henan Normal University, Xinxiang 453007, China}
  \author{Qin Chang}
  \email{changqin@htu.cn}
  \affiliation{Institute of Particle and Nuclear Physics,
              Henan Normal University, Xinxiang 453007, China}
  \author{Jinshu Huang}
  \affiliation{College of Physics and Electronic Engineering,
              Nanyang Normal University, Nanyang 473061, China}
  \author{Yueling Yang}
  \affiliation{Institute of Particle and Nuclear Physics,
              Henan Normal University, Xinxiang 453007, China}

  \begin{abstract}
  Motivated by the recent measurements on nonleptonic $J/{\psi}$ weak
  decays at BESIII and the potential prospects of $J/{\psi}$ meson
  at the high-luminosity heavy-flavor experiments,
  the branching ratios of the two-body nonleptonic $J/{\psi}$ ${\to}$
  $DP$, $DV$ decays are estimated quantitatively by considering the
  QCD radiative corrections to hadronic matrix elements with
  the QCD factorization approach.
  It is found that the Cabibbo favored $J/{\psi}$ ${\to}$ $D_{s}^{-}{\rho}^{+}$,
  $D_{s}^{-}{\pi}^{+}$, $\overline{D}_{u}^{0}\overline{K}^{{\ast}0}$
  decays have branching ratios ${\gtrsim}$ $10^{-10}$,
  which might be promisingly detectable in the near future.
  \end{abstract}
  \keywords{$J/{\psi}$ meson; weak decay; branching ratio; QCD factorization}
  \pacs{13.25.Gv 12.39.St 14.40.Pq 14.65.Dw}
  \maketitle

  \section{Introduction}
  \label{sec01}
  The $J^{PC}$ $=$ $1^{--}$ ground state of the charmonium family,
  the $J/{\psi}$ meson, was discovered in 1974 simultaneously
  both from the $e^{+}e^{-}$ invariant mass by the MIT-BNL group
  \cite{PRL33.1404} and from an enormous increase of the cross
  sections for hadronic, ${\mu}^{+}{\mu}^{-}$ and $e^{+}e^{-}$ final
  states by the SLAC-LBL group \cite{PRL33.1406}.
  Since then the study on the $J/{\psi}$ particle and
  its family members has attracted much persistent attentions
  of experimentalists and theorists due to the facts that:
  on the one hand,
  the charmonium states offer an excellent platform to test and
  improve our understanding of the strong interactions at both
  perturbative and nonperturbative levels;
  on the other hand,
  there is a great renewed interest
  due to the massive dedicated investigation at BES, CLEO-c,
  LHCb and the studies via decays of $B$ mesons at $B$ factories.

  One of the most surprising feature of the $J/{\psi}$ meson is
  its narrow width, ${\Gamma}_{J/{\psi}}$ $=$ $92.9{\pm}2.8$
  keV \cite{pdg},
  which indicates that the decays of $J/{\psi}$ into light
  hadrons are suppressed dynamically.
  The reason for the extremely small decay width of the $J/{\psi}$
  meson is usually referred to by the phenomenological OZI
  (Okubo-Zweig-Iizuka) rules \cite{ozi-o,ozi-z,ozi-i},
  which states that processes with ``detached'' quark lines
  are suppressed.
  It is well known that the mass of the $J/{\psi}$ meson is
  below the $D\bar{D}$ threshold. Hence,
  in despite of the OZI suppression, the $J/{\psi}$ decay
  into hadrons are dominated by the strong and electromagnetic
  interactions, and the decay modes at the lowest order
  approximation could be divided into four types:
  (1) the hadronic decay via the annihilation of the
  $c\bar{c}$ quark pairs into three gluons, {\em i.e.},
  $J/{\psi}$ ${\to}$ $ggg$ ${\to}$ $X$,
  (2) the electromagnetic decay via the $c\bar{c}$
  annihilation into a virtual photon, {\em i.e.},
  $J/{\psi}$ ${\to}$ ${\gamma}^{\ast}$ ${\to}$ $X$,
  (3) the radiative decay via the $c\bar{c}$
  annihilation into one photon and two gluons, {\em i.e.},
  $J/{\psi}$ ${\to}$ ${\gamma}gg$ ${\to}$ ${\gamma}$ $+$ $X$,
  (4) the magnetic dipole transition
  to ${\eta}_{c}$, {\em i.e.},
  $J/{\psi}$ ${\to}$ ${\gamma}{\eta}_{c}$ ${\to}$
  ${\gamma}$ $+$ $X$ \cite{rep174,prog.61.455},
  where $X$ denotes the possible final hadrons.
  Besides, the $J/{\psi}$ meson can decay into hadrons
  also via the weak interactions, although the branching
  ratio for inclusive weak decays via a single $c$ or $\bar{c}$
  quark decay relying on the spectator model is very small,
  about $2/({\tau}_{D}{\Gamma}_{J/{\psi}})$
  ${\sim}$ $10^{-8}$ \cite{pdg,zpc62.271}.
  In this paper, we will concentrate on the flavor-changing
  nonleptonic $J/{\psi}$ ${\to}$ $DM$ weak decays
  with the QCD factorization (QCDF) approach
  \cite{qcdf1,qcdf2,qcdf3,qcdf4,qcdf5,qcdf6},
  where $M$ denotes the low-lying $SU(3)$ pseudoscalar
  and vector meson nonet.
  The reasons are listed as follows.

  From the experimental point of view,
  (1)
  with the running of high-luminosity dedicated heavy-flavor
  factories, more and more $J/{\psi}$ events have been
  accumulating.
  It is hopefully expected to produce about $10^{10}$
  $J/{\psi}$ events at BESIII per year of data taking
  with the designed luminosity \cite{cpc36}, and over
  $10^{10}$ prompt $J/{\psi}$ events at LHCb per $fb^{-1}$
  data \cite{epjc71}.
  The availability of such large samples enables a realistic
  possibility to explore experimentally the nonleptonic
  $J/{\psi}$ weak decays, so the corresponding theoretical
  studies are very necessary to provide a ready reference.
  (2)
  The detection of a single $D$ meson coming from
  the $J/{\psi}$ weak process is free from inefficient
  double tagging of the charmed meson pairs occurring
  above the $D\bar{D}$ threshold.
  In addition, the definite energies and momenta of the
  back-to-back final states in the center-of-mass frame
  of the $J/{\psi}$ meson would provide an unambiguous
  signature.
  With the help of remarkable improvements of experimental
  instrumentation and sophisticated particle identification
  techniques, the accurate measurements on the hadronic
  $J/{\psi}$ ${\to}$ $DM$ weak decays may now be feasible.
  Recently, a search for the Cabibbo favored
  $J/{\psi}$ ${\to}$ $D_{s}{\rho}$, $D_{u}K^{\ast}$ decays
  is performed with available $2.25{\times}10^{8}$
  $J/{\psi}$ events accumulated with the BESIII detector,
  but no evident signal is observed due to
  insufficient statistics \cite{prd89.071101}.
  Of course, such small branching ratios make the observation
  of nonleptonic $J/{\psi}$ weak decays extremely challenging,
  and observation of an abnormally large production
  rate of single charmed mesons in $e^{+}e^{-}$ collisions
  would be a hint of new physics beyond the standard model.

  From the theoretical point of view,
  (1)
  the nonleptonic $J/{\psi}$ weak decay has been studied in
  previous works using the factorization scheme, such as
  Ref.\cite{zpc62.271} based on the spin symmetry
  and nonrecoil approximation,
  Ref.\cite{epjc55} with the QCD sum rules,
  Ref.\cite{prd78} with the covariant light-cone quark model,
  and Refs.\cite{plb252,ijma14,adv2013}
  with the Bauer-Stech-Wirbel (BSW) model \cite{bsw1,bsw2}.
  Due to that the transition form factor is one of the
  essential ingredients for the charmonium weak decay,
  the previous studies \cite{zpc62.271,epjc55,prd78,plb252,ijma14,adv2013}
  mainly concern the calculation of the weak transition
  form factors dominated by the nonperturbative dynamics,
  which lead surely to unavoidable uncertainties on
  theoretical predictions.
  Since the charmonium could be well handled with the nonrelativistic QCD,
  observables of the $J/{\psi}$ ${\to}$ $DM$ decays might be used
  to test and ameliorate various models by comparison with measurements.
  (2)
  In recent years, several attractive QCD-inspired methods have been
  substantially developed and successfully used to cope with the hadronic
  matrix elements of nonleptonic $B$ weak decays, such as
  the soft and collinear effective theory
  \cite{scet1,scet2,scet3,scet4,scet5,scet6,scet7,scet8}
  and QCDF based on the
  collinear factorization approximation and power countering rules
  in the heavy quark limit, the perturbative QCD approach
  \cite{pqcd1,pqcd2,pqcd3,pqcd4,pqcd5,pqcd6}
  based on the $k_{T}$ factorization scheme.
  These methods mainly concern the underlying dynamical mechanism
  of the weak decays of heavy flavor hadrons, and could be
  applied to the weak decays of heavy quarkonium.
  The analysis of nonleptonic $J/{\psi}$ weak decays are particularly
  interesting in exploring mechanism responsible for hadronic transitions
  and very important for study of the applicability of factorization
  theorem and QCD properties at the scale of ${\cal O}(m_{c})$.
  Further, the weak decay of the $J/{\psi}$ particle offers a unique
  opportunity to probe polarization effects involved in vector meson
  decays, which might be helpful to investigate the underlying structure
  and dynamics of heavy quarkonium.

  This paper is organized as follows.
  In section \ref{sec02}, we will present the theoretical framework
  and the amplitudes for nonleptonic $J/{\psi}$ ${\to}$ $DM$
  weak decays within the QCDF framework.
  The section \ref{sec03} is devoted to numerical results and discussion.
  Finally, the section \ref{sec04} is our summation.

  \section{theoretical framework}
  \label{sec02}
  \subsection{The effective Hamiltonian}
  \label{sec0201}
  The low energy effective Hamiltonian responsible for the
  nonleptonic $J/{\psi}$ ${\to}$ $DM$ weak decays
  can be written as \cite{9512380}:
   \begin{equation}
  {\cal H}_{\rm eff}\ =\ \frac{G_{F}}{\sqrt{2}}\,
   \sum\limits_{q_{1},q_{2}}\, V_{cq_{1}}^{\ast} V_{uq_{2}}\,
   \Big\{ C_{1}({\mu})\,Q_{1}({\mu})
         +C_{2}({\mu})\,Q_{2}({\mu}) \Big\}
   + {\rm h.c.}
   \label{hamilton},
   \end{equation}
  where the Fermi coupling constant $G_{F}$ ${\simeq}$
  $1.166{\times}10^{-5}\,{\rm GeV}^{-2}$ \cite{pdg};
  $V_{cq_{1}}^{\ast}V_{uq_{2}}$ is the Cabibbo-Kobayashi-Maskawa
  (CKM) factor and $q_{1,2}$ $=$ $d$, $s$;
  The Wilson coefficients $C_{1,2}(\mu)$ summarize the physical
  contributions above the scale of ${\mu}$.
  The expressions of the local tree four-quark operators are
    \begin{eqnarray}
    Q_{1} &=&
  [ \bar{q}_{1,{\alpha}}{\gamma}_{\mu}(1-{\gamma}_{5})c_{\alpha} ]
  [ \bar{u}_{\beta}{\gamma}^{\mu}(1-{\gamma}_{5})q_{2,{\beta}} ]
    \label{q1}, \\
    Q_{2} &=&
  [ \bar{q}_{1,{\alpha}}{\gamma}_{\mu}(1-{\gamma}_{5})c_{\beta} ]
  [ \bar{u}_{\beta}{\gamma}^{\mu}(1-{\gamma}_{5})q_{2,{\alpha}} ]
    \label{q2},
    \end{eqnarray}
  where ${\alpha}$ and ${\beta}$ are color indices and the
  sum over repeated indices is understood.

  Here, we would like to point out that
  (1)
  due to the large cancellation of the CKM factors
  $V_{cd}^{\ast}V_{ud}$ $+$ $V_{cs}^{\ast}V_{us}$
  ${\sim}$ ${\cal O}({\lambda}^{5})$ where the Wolfenstein
  parameter ${\lambda}$ $=$ ${\sin}{\theta}_{c}$ $=$
  $0.225\,37(61)$ \cite{pdg}
  and ${\theta}_{c}$ is the Cabibbo angle,
  the contributions of penguin and annihilation operators
  are strongly suppressed and could be safely neglected if
  the $CP$-violating asymmetries that are expected to
  be very tiny due to the small weak phase difference for
  $c$ quark decay are prescinded
  from the present consideration.
  (2)
  The Wilson coefficients $C_{i}$ are calculable with the
  perturbation theory and have properly been evaluated to
  the next-to-leading order (NLO).
  Their values at the scale of ${\mu}$ ${\sim}$ ${\cal O}(m_{c})$
  can be obtained with the renormalization group (RG)
  equation \cite{9512380},
  \begin{equation}
  \vec{C}({\mu}) = U_{4}({\mu},m_{b})M(m_{b})U_{5}(m_{b},m_{W})\vec{C}(m_{W})
  \label{ci},
  \end{equation}
  where $U_{f}({\mu}_{f},{\mu}_{i})$ is the RG evolution matrix
  transforming the Wilson coefficients from the scale ${\mu}_{i}$
  to ${\mu}_{f}$,
  and $M({\mu})$ is the quark threshold matching matrix.
  The explicit expressions of $U_{f}({\mu}_{f},{\mu}_{i})$ and
  $M({\mu})$ can be found in Ref.\cite{9512380}.
  The numerical values of LO and NLO $C_{1,2}$ in naive
  dimensional regularization scheme
  are listed in Table \ref{tab01}.
  The values of NLO Wilson coefficients in Table
  \ref{tab01} are consistent with those
  given by Refs.\cite{bsw2,9512380,ijtp53},
  where a trick with ``effective'' number of active flavors
  $f$ $=$ $4.15$ rather than formula Eq.(\ref{ci}) is used
  by Ref.\cite{9512380}.
  (3)
  To obtain the decay amplitudes and branching ratios,
  the remaining works are how to accurately evaluate the
  hadronic matrix elements where the local operators are
  sandwiched between the initial and final states,
  which is also the most intricate melody in dealing
  with the weak decay of heavy hadrons by now.

  \subsection{Hadronic matrix elements}
  \label{sec0202}
  Phenomenologically, the simplest treatment on hadronic matrix elements
  of a four fermion operator is the approximation by the product of the
  decay constants and the transition form factors based on the color
  transparency ansatz \cite{bjorken} and the naive factorization
  scheme (NF) \cite{nf1,nf2}.
  As well known, the NF's defects are very obvious and displayed
  as the absence of the renormalization scale dependence, the strong
  phases and the nonfactorizatable corrections from the hadronic
  matrix elements, which result in nonphysical decay amplitudes and
  the incapacity of prediction on $CP$-violating asymmetries.
  To remedy this situation, M. Beneke {\em et al.} \cite{qcdf1,qcdf2} proposed that
  the hadronic matrix elements could be written as the convolution
  integrals of hard scattering kernels and the light cone distribution
  amplitudes with the QCDF approach.

  Using the QCDF master formula, the hadronic matrix elements for
  the $J/{\psi}$ ${\to}$ $DM$ decays could be expressed as :
   \begin{equation}
  {\langle}DM{\vert}Q_{i}{\vert}J/{\psi}{\rangle}
   \ =\
   \sum\limits_{i} F_{i}^{ J{\to}D }
  {\int}\,dx\, H_{i}(x)\,{\Phi}_{M}(x)
   \ =\
   \sum\limits_{i} F_{i}^{ J{\to}D }f_{M}
   \{1+\frac{{\alpha}_{s}}{\pi}r+{\cdots}\}
   \label{hadronic},
   \end{equation}
  where $F_{i}^{ J{\to}D }$ is the transition form factor and
  ${\Phi}_{M}(x)$ is the light cone distribution amplitude of
  the emitted meson $M$ with the decay constant $f_{M}$,
  which are assumed to be dominated by nonperturbative contributions
  and taken as universal inputs.

  Here, we would like to point out that
  (1)
  for the $J/{\psi}$ ${\to}$ $DM$ decay,
  the spectator quark is the almost on-shell charm (anti)quark.
  It is commonly thought that the virtuality of the gluon tied up
  with the heavy spectator quark is of order
  ${\Lambda}_{\rm QCD}^{2}$.
  The hard and soft contributions associated with the charmed
  spectator entangle with each other and
  cannot be separated properly.
  According to the basic idea of the QCDF approach \cite{qcdf2},
  the physical form factors that could be obtained from lattice
  QCD or QCD sum rules are introduced as inputs, and the hard
  spectator scattering contributions that are power suppressed
  in the heavy quark limit disappeared from Eq.(\ref{hadronic}).
  (2)
  The hard scattering kernels $H_{i}(x)$, including the
  nonfactorizable vertex corrections, are computable order by
  order with the perturbation theory in principle.
  At the order ${\alpha}_{s}^{0}$,
  $H_{i}(x)$ $=$ $1$, {\em i.e.}, the convolution integral of
  Eq.(\ref{hadronic}) results in a decay constant
  and one goes back to the simple NF scenario.
  At the order ${\alpha}_{s}$ and higher orders,
  the information of the renormalization scale dependence
  and strong phases hidden in hadronic matrix elements
  could be partly recuperated.
  Combined the nonfactorizable contributions with the
  Wilson coefficients, the scale independent effective
  coefficients at the order ${\alpha}_{s}$ can be obtained
  \cite{qcdf3}:
  \begin{eqnarray}
   a_{1}
   &=& C_{1}^{\rm NLO}+\frac{1}{N_{c}}\,C_{2}^{\rm NLO}
    + \frac{{\alpha}_{s}}{4{\pi}}\, \frac{C_{F}}{N_{c}}\,
      C_{2}^{\rm LO}\, V
   \label{a1}, \\
   a_{2}
   &=& C_{2}^{\rm NLO}+\frac{1}{N_{c}}\,C_{1}^{\rm NLO}
    + \frac{{\alpha}_{s}}{4{\pi}}\, \frac{C_{F}}{N_{c}}\,
      C_{1}^{\rm LO}\, V
   \label{a2}.
  \end{eqnarray}

  The expression of vertex corrections could be written
  as \cite{qcdf3}:
  \begin{equation}
  V = 6\,{\log}\Big( \frac{m_{c}^{2}}{{\mu}^{2}} \Big)
    -  18 - \Big( \frac{1}{2}+i3{\pi} \Big)\,a_{0}^{M}
    +  \Big( \frac{11}{2}-i3{\pi} \Big)\,a_{1}^{M}
    -   \frac{21}{20}\,a_{2}^{M} +{\cdots}
  \label{vc},
  \end{equation}
  with the twist-2 distribution amplitudes of
  pseudoscalar and longitudinally polarized vector meson
  in terms of Gegenbauer polynomials \cite{ball1,ball2,ball3}:
   \begin{equation}
  {\phi}_{M}(x)=6\,x\bar{x}
   \sum\limits_{n=0}^{\infty}
   a_{n}^{M}\, C_{n}^{3/2}(x-\bar{x})
   \label{twist},
   \end{equation}
  where $\bar{x}$ $=$ $1$ $-$ $x$;
  $a_{n}^{M}$ is the Gegenbauer moment
  and $a_{0}^{M}$ ${\equiv}$ $1$.

  It is found that
  (1) for the coefficient $a_{1}$, nonfactorizable vertex
  corrections can provide ${\ge}$ 10\% enhancement compared
  with the NF's value, and a small strong phase ${\le}$ $5^{\circ}$.
  (2) for the coefficient $a_{2}$, contributions of
  vertex corrections assisted with the large Wilson coefficient
  $C_{1}$ are significant, and a relatively large strong
  phase ${\sim}$ $-115^{\circ}$ is obtained.
  (3) the magnitude of $a_{1,2}$ agrees well with that
  from the fit on hadronic $D$ weak decays \cite{ai1,ai2}, but with
  more information on the strong phases.

  \subsection{Decay amplitude}
  \label{sec0203}
  Within the QCDF framework, the Lorentz-invariant amplitudes
  for $J/{\psi}$ ${\to}$ $DM$ decays can be expressed as:
   \begin{equation}
  {\cal A}(J/{\psi}{\to}DM)\ =\
  {\langle}DM{\vert}{\cal H}_{\rm eff}{\vert}J/{\psi}{\rangle}\ =\
   \frac{G_{F}}{\sqrt{2}}\,
   V_{cq_{1}}^{\ast} V_{uq_{2}}\, a_{i}\,
  {\langle}M{\vert}J^{\mu}{\vert}0{\rangle}
  {\langle}D{\vert}J_{\mu}{\vert}J/{\psi}{\rangle}
   \label{lorentz}.
   \end{equation}

  The matrix elements of current operators are defined as follows:
   \begin{eqnarray}
  {\langle}P(p){\vert}A_{\mu}{\vert}0{\rangle}
  &=&
   -if_{P}\,p_{\mu}
   \label{cme01}, \\
  {\langle}V(p,{\epsilon}){\vert}V_{\mu}{\vert}0{\rangle}
  &=&
   f_{V}\,m_{V}\,{\epsilon}_{\mu}^{\ast}
   \label{cme02},
   \end{eqnarray}
 where $f_{P}$ and $f_{V}$ are the decay constants
 of pseudoscalar and vector mesons, respectively;
 $m_{V}$ and ${\epsilon}$ denote the mass and
 polarization of vector meson, respectively.

  The transition form factors are defined as follows
  \cite{epjc55,prd78,plb252,ijma14,adv2013,bsw1,bsw2}:
    \begin{eqnarray}
   & &
   {\langle}D(p_{2}){\vert}V_{\mu}-A_{\mu}
   {\vert}J/{\psi}(p_{1},{\epsilon}){\rangle}
    \nonumber \\ &=&
  -{\epsilon}_{{\mu}{\nu}{\alpha}{\beta}}\,
   {\epsilon}_{J}^{{\nu}}\,
    q^{\alpha}\, (p_{1}+p_{2})^{\beta}\,
     \frac{V^{J{\to}D}(q^{2})}{m_{J}+m_{D}}
   -i\,\frac{2\,m_{J}\,{\epsilon}_{J}{\cdot}q}{q^{2}}\,
    q_{\mu}\, A_{0}^{J{\to}D}(q^{2})
    \nonumber \\ & &
    -i\,{\epsilon}_{J,{\mu}}\,
    ( m_{J}+m_{D} )\, A_{1}^{J{\to}D}(q^{2})
   -i\,\frac{{\epsilon}_{J}{\cdot}q}{m_{J}+m_{D}}\,
   ( p_{1} + p_{2} )_{\mu}\, A_{2}^{J{\to}D}(q^{2})
    \nonumber \\ & &
   +i\,\frac{2\,m_{J}\,{\epsilon}_{J}{\cdot}q}{q^{2}}\,
   q_{\mu}\, A_{3}^{J{\to}D}(q^{2})
    \label{cme03},
    \end{eqnarray}
  where $q$ $=$ $p_{1}$ $-$ $p_{2}$;
  and $A_{0}(0)$ $=$ $A_{3}(0)$
  is required compulsorily to cancel singularities at the
  pole $q^{2}$ $=$ $0$.
  There is a relation among these form factors
  \begin{equation}
  2m_{J}A_{3}(q^{2})=(m_{J}+m_{D})A_{1}(q^{2})+(m_{J}-m_{D})A_{2}(q^{2})
  \label{form01}.
  \end{equation}

  It is clearly seen that
  there are only three independent form factors, $A_{0,1}(0)$
  and $V(0)$, at the pole $q^{2}$ $=$ $0$ for the hadronic
  $J/{\psi}$ ${\to}$ $DM$ decays.
  From the convolution integral expressions of form factors
  at zero momentum transfer in terms of participating
  meson wave functions given in Refs.\cite{adv2013,bsw1,bsw2},
  there is approximately a hierarchic relationship, {\em i.e.},
  \begin{equation}
  V^{J{\to}D}(0)\,{\approx}\,3A_{1}^{J{\to}D}(0),
  \quad
  A_{1}^{J{\to}D}(0)\,{\ge}\,A_{0}^{J{\to}D}(0)
  \label{form02},
  \end{equation}
  which are also verified by the numbers of Table 1 in Ref.\cite{adv2013}.

  With the above definition, amplitudes for $J/{\psi}$ ${\to}$ $DP$, $DV$
  decay are explicitly listed in the Appendix \ref{app01} and \ref{app02}.
  Here, we would like to point out that
  (1)
  the amplitudes for $J/{\psi}$ ${\to}$ $DV$ decays are
  conventionally expressed by helicity amplitudes, which
  They are defined by the decomposition \cite{vv1,vv2,vv3},
    \begin{eqnarray}
   {\cal H}_{\lambda} & =&
   {\langle}V{\vert}J^{\mu}{\vert}0{\rangle}
   {\langle}D{\vert}J_{\mu}{\vert}J/{\psi}{\rangle}
    \nonumber \\ &=&
   {\epsilon}_{V}^{{\ast}{\mu}} {\epsilon}_{J}^{\nu} \Big\{
    a\,g_{{\mu}{\nu}}
    +\frac{ b  }{m_{J}\,m_{V}} ( p_{J} + p_{D} )^{\mu} p_{V}^{\nu}
    +\frac{i\,c }{m_{J}\,m_{V}} {\epsilon}_{{\mu}{\nu}{\alpha}{\beta}}
     p_{V}^{\alpha}(p_{J}+p_{D})^{\beta} \Big\}
    \label{spd}.
    \end{eqnarray}
 The relations among helicity amplitudes and invariant amplitudes $a$, $b$, $c$ are
    \begin{eqnarray}
   {\cal H}_{0} &=& -a\,x-2b\,(x^{2}-1)
    \label{h0}, \\
   {\cal H}_{\pm} &=& a\, {\pm}\,2c\,\sqrt{x^{2}-1}
    \label{h1},
    \end{eqnarray}
  where the expressions of $a$, $b$, $c$ and $x$ are
    \begin{eqnarray}
    a &=& -i\, f_{V}\, m_{V}\, ( m_{J}+m_{D} )\, A_{1}^{J{\to}D}(q^{2})
    \label{sa}, \\
    b &=& -i\, f_{V}\, m_{J}\, m_{V}^{2}\, \frac{A_{2}^{J{\to}D}(q^{2})}{m_{J}+m_{D}}
    \label{db}, \\
   c &=& +i\, f_{V}\, m_{J}\, m_{V}^{2}\, \frac{V^{J{\to}D}(q^{2})}{m_{J}+m_{D}}
    \label{pc}, \\
    x &=& \frac{p_{J}{\cdot}p_{V}}{m_{J}\,m_{V}}
      \ =\ \frac{m_{J}^{2}-m_{D}^{2}+m_{V}^{2}}{2\,m_{J}\,m_{V}}
    \label{xx}.
    \end{eqnarray}
  There scalar amplitudes $a$, $b$, $c$ describe the $s$, $d$, $p$
  wave contributions, respectively.
  Clearly, compared with the $s$ wave amplitude,
  the $p$ and $d$ wave amplitudes are suppressed
  by a factor of $m_{V}/m_{J}$.
  (2)
  The light vector mesons are assumed ideally mixed,
  {\em i.e.}, the ${\omega}$ $=$ $(u\bar{u}+d\bar{d})/\sqrt{2}$
  and ${\phi}$ $=$ $s\bar{s}$.
  As for the mixing of pseudoscalar ${\eta}$ and ${\eta}^{\prime}$
  meson, we will adopt the quark-flavor basis description proposed
  in Ref.\cite{eta}, and neglect the contributions from
  possible gluonium and $c\bar{c}$ compositions, {\em i.e.},
   \begin{equation}
   \left(\begin{array}{c}
  {\eta} \\ {\eta}^{\prime}
   \end{array}\right) =
   \left(\begin{array}{cc}
  {\cos}{\phi} & -{\sin}{\phi} \\
  {\sin}{\phi} &  {\cos}{\phi}
   \end{array}\right)
   \left(\begin{array}{c}
  {\eta}_{q} \\ {\eta}_{s}
   \end{array}\right)
   \label{mixing01},
   \end{equation}
  where ${\eta}_{q}$ $=$ $(u\bar{u}+d\bar{d})/{\sqrt{2}}$
  and ${\eta}_{s}$ $=$ $s\bar{s}$, respectively;
  the mixing angle ${\phi}$ $=$ $(39.3{\pm}1.0)^{\circ}$
  \cite{eta}.
  The mass relations between physical states (${\eta}$
  and ${\eta}^{\prime}$) and flavor states (${\eta}_{q}$
  and ${\eta}_{s}$) are
   \begin{eqnarray}
   m_{{\eta}_{q}}^{2}&=& \displaystyle
   m_{\eta}^{2}{\cos}^{2}{\phi}
  +m_{{\eta}^{\prime}}^{2}{\sin}^{2}{\phi}
  -\frac{\sqrt{2}f_{{\eta}_{s}}}{f_{{\eta}_{q}}}
  (m_{{\eta}^{\prime}}^{2}- m_{\eta}^{2})\,
  {\cos}{\phi}\,{\sin}{\phi}
   \label{ss12}, \\
   m_{{\eta}_{s}}^{2}&=& \displaystyle
   m_{\eta}^{2}{\sin}^{2}{\phi}
  +m_{{\eta}^{\prime}}^{2}{\cos}^{2}{\phi}
  -\frac{f_{{\eta}_{q}}}{\sqrt{2}f_{{\eta}_{s}}}
  (m_{{\eta}^{\prime}}^{2}- m_{\eta}^{2})\,
  {\cos}{\phi}\,{\sin}{\phi}
   \label{ss13}.
   \end{eqnarray}

  \section{Numerical results and discussion}
  \label{sec03}

  In the rest frame of $J/{\psi}$ particle, branching ratio
  for nonleptonic $J/{\psi}$ weak decays can be written as
   \begin{equation}
  {\cal B}(J/{\psi}{\to}DM)\ =\ \frac{1}{12{\pi}}\,
   \frac{p_{\rm cm}}{m_{J}^{2}{\Gamma}_{J}}\,
  {\vert}{\cal A}(J/{\psi}{\to}DM){\vert}^{2}
   \label{br},
   \end{equation}
 where the common momentum of final states is
   \begin{equation}
   p_{\rm cm}\ =\
   \frac{ \sqrt{ [m_{J}^{2}-(m_{D}+m_{M})^{2}]  [m_{J}^{2}-(m_{D}-m_{M})^{2}] }  }{2\,m_{J}}
   \label{pcm}.
   \end{equation}

 The input parameters in our calculation,
 including the CKM Wolfenstein parameters,
 decay constants of mesons, Gegenbauer moments of distribution
 amplitudes in Eq.(\ref{twist}), are collected in Table \ref{tab02}.
 If not specified explicitly, we will take their central
 values as the default inputs.
 As well known, the transition form factors are essential parameters
 in the QCDF master formula of Eq.(\ref{hadronic}),
 but the discrepancy among previous results on form factors with
 different models (see Table 1 of Ref.\cite{adv2013}) is still large.
 In this paper, we will use the mean values of the form factors
 given in Ref.\cite{plb252} with additional uncertainties
 to offer an order of magnitude estimation, {\em i.e.},
   \begin{eqnarray}
   & &
   A_{0}^{J{\to}D}(0)\ =\ 0.50{\pm}0.1,
   \quad
   A_{0}^{J{\to}D_{s}}(0)\ =\ 0.55{\pm}0.1
   \label{ffa0}, \\
   & &
   A_{1}^{J{\to}D}(0)\ =\ 0.55{\pm}0.1,
   \quad
   A_{1}^{J{\to}D_{s}}(0)\ =\ 0.65{\pm}0.1
   \label{ffa1}, \\
   & &
   V^{J{\to}D}(0)\ =\ 1.50{\pm}0.3,
   \quad
   V^{J{\to}D_{s}}(0)\ =\ 1.50{\pm}0.3
   \label{ffv}.
   \end{eqnarray}

 Our numerical results on the $CP$-averaged branching ratios
 for $J/{\psi}$ ${\to}$ $DP$, $DV$ decays are displayed in
 Table \ref{tab03}, where theoretical uncertainties of the
 last column come from the CKM parameters,
 the renormalization scale ${\mu}$ $=$ $(1{\pm}0.2)m_{c}$,
 decay constants and Gegenbauer moments,
 transition form factors, respectively.
 For comparison, the previous results \cite{epjc55,ijma14,adv2013}
 with coefficients $a_{1}$ $=$ $1.26$ and $a_{2}$ $=$ $-0.51$
 are also listed in the columns 3--7 of Table \ref{tab03},
 where numbers in the columns 3 and 4--7 are calculated
 with different form factors based on QCD sum rules and
 BSW model, respectively; numbers in the columns 4--7
 correspond to the form factors given by
 the flavor dependent ${\omega}$ (``A'' column),
 QCD inspired ${\omega}$ $=$ ${\alpha}_{s}{\times}m$ (``B'' column),
 the universal ${\omega}$ $=$ 4 GeV (``C'' column)
 and ${\omega}$ $=$ 5 GeV (``D'' column), respectively;
 and ${\omega}$, the average transverse quark momentum,
 is a parameter of the BSW wave functions.
 The following are some comments.

 (1)
 There are some differences among the estimations (see the numbers
 in Table \ref{tab03}) on branching ratios for hadronic $J/{\psi}$
 ${\to}$ $DP$, $DV$ decays.
 (i)
 The discrepancies among previous works, although the same values
 of coefficients $a_{1,2}$ are taken, come mainly from different values
 of form factors.
 (ii)
 Considering the effects of nonfactorizable vertex corrections,
 the QCDF's predictions on branching ratios agree basically with
 previous results, at least with the same order magnitude.
 The QCDF's results are generally in line with the numbers in
 columns ``C'' and ``D'' within uncertainties, because the values
 of form factors in our calculation is close to the average values
 of form factors used in columns ``C'' and ``D''.

 (2) There are some hierarchical structures.
 (i)
 According to the coefficients $a_{1,2}$ and the CKM factors,
 the $J/{\psi}$ ${\to}$ $DP$, $DV$ decays could be divided
 into different cases listed below.
 \begin{center}
 \begin{ruledtabular}
 \begin{tabular}{ccccc}
 Case & Coefficient & CKM factor & Branching ratio & Decay modes\\ \hline
 1 a  & $a_{1}$ & ${\vert}V_{ud}V_{cs}^{\ast}{\vert}$ ${\sim}$ $1$
                & ${\gtrsim}$ $10^{- 9}$ & $D_{s}{\rho}$ \\
              & & & ${\gtrsim}$ $10^{-10}$ & $D_{s}{\pi}$ \\
 1 b  & $a_{1}$ & ${\vert}V_{ud}V_{cd}^{\ast}{\vert}$,
                  ${\vert}V_{us}V_{cs}^{\ast}{\vert}$ ${\sim}$ ${\lambda}$
                & ${\gtrsim}$ $10^{-10}$ & $D_{s}K^{\ast}$, $D_{d}{\rho}$ \\
              & & & ${\gtrsim}$ $10^{-11}$ & $D_{s}K$, $D_{d}{\pi}$  \\
 1 c  & $a_{1}$ & ${\vert}V_{us}V_{cd}^{\ast}{\vert}$ ${\sim}$ ${\lambda}^{2}$
                & ${\gtrsim}$ $10^{-12}$ & $D_{d}K^{\ast}$, $D_{d}K$ \\ \hline
 2 a  & $a_{2}$ & ${\vert}V_{ud}V_{cs}^{\ast}{\vert}$ ${\sim}$ $1$
                & ${\gtrsim}$ $10^{-10}$ & $D_{u}K^{\ast}$ \\
              & & & ${\gtrsim}$ $10^{-11}$ & $D_{u}K$ \\
 2 b  & $a_{2}$ & ${\vert}V_{ud}V_{cd}^{\ast}{\vert}$,
                  ${\vert}V_{us}V_{cs}^{\ast}{\vert}$ ${\sim}$ ${\lambda}$
                & ${\gtrsim}$ $10^{-11}$ & $D_{u}{\phi}$ \\
              & & & ${\gtrsim}$ $10^{-12}$
                  & $D_{u}{\rho}$, $D_{u}{\omega}$, $D_{u}{\pi}$, $D_{u}{\eta}$  \\
 2 c  & $a_{2}$ & ${\vert}V_{us}V_{cd}^{\ast}{\vert}$ ${\sim}$ ${\lambda}^{2}$
                   & ${\gtrsim}$ $10^{-13}$ & $D_{u}\overline{K}^{\ast}$, $D_{u}\overline{K}$
 \end{tabular}
 \end{ruledtabular}
 \end{center}
 The extremely small branching ratios for $J/{\psi}$ ${\to}$
 $D_{u}{\eta}^{\prime}$ decays is mainly due to
 the cancellation of CKM factors between
 $V_{ud}V_{cd}^{\ast}$ ${\sim}$ $-{\lambda}$ and
 $V_{us}V_{cs}^{\ast}$ ${\sim}$ $+{\lambda}$
 resulting in the destructive interferences between
 the amplitudes ${\cal A}(J/{\psi}{\to}D_{u}{\eta}_{q})$
 and ${\cal A}(J/{\psi}{\to}D_{u}{\eta}_{s})$.
 (ii)
 Compared with the $J/{\psi}$ ${\to}$ $DV$ decays,
 the $J/{\psi}$ ${\to}$ $DP$ decays are suppressed
 dynamically by the orbital angular momentum
 of final states $L_{DP}$ $>$ $L_{DV}$.
 (iii)
 In addition, after a careful scrutiny of the QCDF's results,
 it is interestingly found that there is an approximative
 relationship among the branching ratios for decay modes
 with the same charmed final state,
 ${\cal B}(J/{\psi}{\to}DV)$ ${\approx}$
 $5\,{\cal B}(J/{\psi}{\to}DP)$, where the pseudoscalar $P$
 and vector $V$ mesons corresponds to each other in the
 $SU(3)$ $q\bar{q}$ assignments of light mesons with the
 quark model, such as,
 ${\cal B}(J/{\psi}{\to}D_{s}^{-}{\rho}^{+})$ ${\approx}$
 $5\,{\cal B}(J/{\psi}{\to}D_{s}^{-}{\pi}^{+})$ and so on.
 Hence, the CKM-favored $a_{1}$ dominated $J/{\psi}$
 ${\to}$ $D_{s}^{-}{\rho}^{+}$ decay has the largest
 branching ratio, which should be sought for with high
 priority and firstly observed by experimental physicists.

 (3)
 There are many uncertainties on the QCDF's results.
 (i)
 The first uncertainty from the CKM factors is small due to
 the high precision on Wolfenstein parameter ${\lambda}$
 with only 0.3\% relative errors now\cite{pdg}.
 The second uncertainty from the renormalization scale
 could, in principle, be reduced by the inclusion of higher
 order ${\alpha}_{s}$ corrections to hadronic matrix elements,
 for example, it has been showed \cite{nnlo1,nnlo2} that tree amplitudes
 incorporating with the NNLO vertex corrections are relatively
 less sensitive to the choice of scale than the NLO amplitudes.
 The largest uncertainty (the fourth uncertainty), ${\sim}$ $40\%$,
 comes from the transition form factors, which is expected
 to be cancelled from the relative ratio of branching ratios,
 such as,
   \begin{eqnarray}
    R_{1} &=&
   \frac{ {\cal B}(J/{\psi}{\to}D_{s}^{-}K^{+}) }{ {\cal B}(J/{\psi}{\to}D_{s}^{-}{\pi}^{+}) }
   \ {\approx}\ {\vert}V_{us}{\vert}^{2}\frac{ f_{K}^{2} }{ f_{\pi}^{2} }
   \ {\approx}\ ( 5.66^{+0.03+0.01+0.10+0.00}_{-0.03-0.00-0.10-0.00}) \%
   \label{r01}, \\
    R_{2} &=&
   \frac{ {\cal B}(J/{\psi}{\to}D_{d}^{-}K^{+}) }{ {\cal B}(J/{\psi}{\to}D_{d}^{-}{\pi}^{+}) }
   \ {\approx}\ {\vert}V_{us}{\vert}^{2}\frac{ f_{K}^{2} }{ f_{\pi}^{2} }
   \ {\approx}\ ( 5.95^{+0.03+0.01+0.11+0.00}_{-0.03-0.00-0.10-0.00}) \%
   \label{r02}.
   \end{eqnarray}
 (ii)
 Uncertainties from other factors, such as the contributions
 of higher order ${\alpha}_{s}$ corrections to hadronic matrix elements,
 $q^{2}$ dependence of form factors,
 the final state interactions and so on,
 which deserve the dedicated study, are not considered in this paper.
 So one should not be too serious about the absolute size of the
 QCDF's branching ratios for $J/{\psi}$ ${\to}$ $DM$ decays
 which just provide an order of magnitude estimation.

  \section{Summary}
  \label{sec04}
  In this paper, we present a phenomenological study on
  the nonleptonic $J/{\psi}$ ${\to}$ $DP$,
  $DV$ weak decays with the QCDF approach.
  Our attention was fixed on the nonfactorizable contributions to hadronic
  matrix elements, while the weak transition form factors are taken as
  nonperturbative parameters,  which is different from previous works
  \cite{zpc62.271,epjc55,prd78,plb252,ijma14,adv2013}.
  The values of coefficients $a_{1,2}$ incorporating QCD
  radiative corrections agree well with those obtained from the fit on hadronic
  $D$ weak decays \cite{ai1,ai2}, which imply that the QCDF approach might
  be valid for the $J/{\psi}$ weak decays.
  Then the branching ratios of the exclusive $J/{\psi}$ ${\to}$ $DP$,
  $DV$ weak decays are estiamted.
  It is found that the QCDF's predictions on branching ratios are rough
  due to large uncertainties from input parameters,  especially from
  form factors.
  Despite this, we still can get some information about the $J/{\psi}$
  ${\to}$ $DP$, $DV$ decays.
  For example, the Cabibbo favored $J/{\psi}$ ${\to}$ $D_{s}^{-}{\rho}^{+}$,
  $D_{s}^{-}{\pi}^{+}$, $\overline{D}_{u}^{0}\overline{K}^{{\ast}0}$
  decays have relatively large branching ratios compared with other
  decay modes, which are promisingly detected at the high-luminosity
  heavy-flavor experiments in the forthcoming years.

  \section*{Acknowledgments}
  The work is supported by both the National Natural Science Foundation
  of China (Grant Nos. 11475055, 11275057, U1232101 and U1332103)
  and the Program for Science and Technology Innovation Talents in
  Universities of Henan Province (Grant No. 2012HASTIT030 and 14HASTIT036).
  Q. Chang is also supported by the Foundation for the Author of
  National Excellent Doctoral Dissertation of China (Grant No. 201317).

  \begin{appendix}
  \section{The amplitudes for $J/{\psi}$ ${\to}$ $DP$ decays}
  \label{app01}
  \begin{eqnarray}
  {\cal A}(J/{\psi}{\to}D_{s}^{-}{\pi}^{+})
  &=&
   \sqrt{2}\, G_{F}\, m_{J}\, ({\epsilon}_{J}{\cdot}p_{\pi})\,
   f_{\pi}\, A_{0}^{J{\to}D_{s}}\,
   V_{cs}^{\ast}\, V_{ud}\, a_{1}
   \label{amp-cs-pi}, \\
  {\cal A}(J/{\psi}{\to}D_{s}^{-}K^{+})
  &=&
   \sqrt{2}\, G_{F}\, m_{J}\, ({\epsilon}_{J}{\cdot}p_{K})\,
   f_{K}\, A_{0}^{J{\to}D_{s}}\,
   V_{cs}^{\ast}\, V_{us}\, a_{1}
   \label{amp-cs-k}, \\
  {\cal A}(J/{\psi}{\to}D_{d}^{-}{\pi}^{+})
  &=&
   \sqrt{2}\, G_{F}\, m_{J}\, ({\epsilon}_{J}{\cdot}p_{\pi})\,
   f_{\pi}\, A_{0}^{J{\to}D_{d}}\,
   V_{cd}^{\ast}\, V_{ud}\, a_{1}
   \label{amp-cd-pi}, \\
  {\cal A}(J/{\psi}{\to}D_{d}^{-}K^{+})
  &=&
   \sqrt{2}\, G_{F}\, m_{J}\, ({\epsilon}_{J}{\cdot}p_{K})\,
   f_{K}\, A_{0}^{J{\to}D_{d}}\,
   V_{cd}^{\ast}\, V_{us}\, a_{1}
   \label{amp-cd-k}, \\
  {\cal A}(J/{\psi}{\to}\overline{D}_{u}^{0}{\pi}^{0})
  &=&
  -G_{F}\, m_{J}\, ({\epsilon}_{J}{\cdot}p_{\pi})\,
   f_{\pi}\, A_{0}^{J{\to}D_{u}}\,
   V_{cd}^{\ast}\, V_{ud}\, a_{2}
   \label{amp-cu-pi}, \\
  {\cal A}(J/{\psi}{\to}\overline{D}_{u}^{0}K^{0})
  &=&
   \sqrt{2}\, G_{F}\, m_{J}\, ({\epsilon}_{J}{\cdot}p_{K})\,
   f_{K}\, A_{0}^{J{\to}D_{u}}\,
   V_{cd}^{\ast}\, V_{us}\, a_{2}
   \label{amp-cu-kz}, \\
  {\cal A}(J/{\psi}{\to}\overline{D}_{u}^{0}\overline{K}^{0})
  &=&
   \sqrt{2}\, G_{F}\, m_{J}\, ({\epsilon}_{J}{\cdot}p_{K})\,
   f_{K}\, A_{0}^{J{\to}D_{u}}\,
   V_{cs}^{\ast}\, V_{ud}\, a_{2}
   \label{amp-cu-kzb}, \\
  {\cal A}(J/{\psi}{\to}\overline{D}_{u}^{0}{\eta}_{q})
  &=&
   G_{F}\, m_{J}\, ({\epsilon}_{J}{\cdot}p_{{\eta}_{q}})\,
   f_{{\eta}_{q}}\, A_{0}^{J{\to}D_{u}}\,
   V_{cd}^{\ast}\, V_{ud}\, a_{2}
   \label{amp-cu-etaq}, \\
  {\cal A}(J/{\psi}{\to}\overline{D}_{u}^{0}{\eta}_{s})
  &=&
   \sqrt{2}\, G_{F}\, m_{J}\, ({\epsilon}_{J}{\cdot}p_{{\eta}_{s}})\,
   f_{{\eta}_{s}}\, A_{0}^{J{\to}D_{u}}\,
   V_{cs}^{\ast}\, V_{us}\, a_{2}
   \label{amp-cu-etas}, \\
  {\cal A}(J/{\psi}{\to}\overline{D}_{u}^{0}{\eta})
  &=&
  {\cos}{\phi}\,{\cal A}(J/{\psi}{\to}\overline{D}_{u}^{0}{\eta}_{q})
 -{\sin}{\phi}\,{\cal A}(J/{\psi}{\to}\overline{D}_{u}^{0}{\eta}_{s})
   \label{amp-cu-eta}, \\
  {\cal A}(J/{\psi}{\to}\overline{D}_{u}^{0}{\eta}^{\prime})
  &=&
  {\sin}{\phi}\,{\cal A}(J/{\psi}{\to}\overline{D}_{u}^{0}{\eta}_{q})
 +{\cos}{\phi}\,{\cal A}(J/{\psi}{\to}\overline{D}_{u}^{0}{\eta}_{s})
   \label{amp-cu-etap}.
  \end{eqnarray}

  \section{The amplitudes for $J/{\psi}$ ${\to}$ $DV$ decays}
  \label{app02}
  \begin{eqnarray}
  {\cal A}(J/{\psi}{\to}D_{s}^{-}{\rho}^{+})
  &=&
   -i\,\frac{G_{F}}{\sqrt{2}}\, f_{\rho}\, m_{\rho}\,
   V_{cs}^{\ast}\, V_{ud}\, a_{1}\, \Big\{
   ({\epsilon}_{\rho}^{\ast}{\cdot}{\epsilon}_{J})\,
   (m_{J}+m_{D_{s}})\, A_{1}^{J{\to}D_{s}}
   \nonumber \\ & & \hspace{-15mm}
   + ({\epsilon}_{\rho}^{\ast}{\cdot}p_{J})\,
  ({\epsilon}_{J}{\cdot}p_{\rho})\,
   \frac{2\, A_{2}^{J{\to}D_{s}}}{ m_{J}+m_{D_{s}} }
  -i\,{\epsilon}_{{\mu}{\nu}{\alpha}{\beta}}\,
  {\epsilon}_{\rho}^{{\ast}{\mu}}\,{\epsilon}_{J}^{\nu}\,
  p_{\rho}^{\alpha}\,p_{J}^{\beta}\,
  \frac{2\, V^{J{\to}D_{s}}}{ m_{J}+m_{D_{s}} }
   \Big\}
   \label{amp-cs-rho}, \\
  {\cal A}(J/{\psi}{\to}D_{s}^{-}K^{{\ast}+})
  &=&
   -i\,\frac{G_{F}}{\sqrt{2}}\, f_{K^{\ast}}\, m_{K^{\ast}}\,
   V_{cs}^{\ast}\, V_{us}\, a_{1}\, \Big\{
   ({\epsilon}_{K^{\ast}}^{\ast}{\cdot}{\epsilon}_{J})\,
   (m_{J}+m_{D_{s}})\, A_{1}^{J{\to}D_{s}}
   \nonumber \\ & & \hspace{-15mm}
   + ({\epsilon}_{K^{\ast}}^{\ast}{\cdot}p_{J})\,
  ({\epsilon}_{J}{\cdot}p_{K^{\ast}})\,
   \frac{2\, A_{2}^{J{\to}D_{s}}}{ m_{J}+m_{D_{s}} }
  -i\,{\epsilon}_{{\mu}{\nu}{\alpha}{\beta}}\,
  {\epsilon}_{K^{\ast}}^{{\ast}{\mu}}\,{\epsilon}_{J}^{\nu}\,
  p_{K^{\ast}}^{\alpha}\,p_{J}^{\beta}\,
  \frac{2\, V^{J{\to}D_{s}}}{ m_{J}+m_{D_{s}} }
   \Big\}
   \label{amp-cs-kv}, \\
  {\cal A}(J/{\psi}{\to}D_{d}^{-}{\rho}^{+})
  &=&
  -i\,\frac{G_{F}}{\sqrt{2}}\, f_{\rho}\, m_{\rho}\,
   V_{cd}^{\ast}\, V_{ud}\, a_{1}\, \Big\{
   ({\epsilon}_{\rho}^{\ast}{\cdot}{\epsilon}_{J})\,
   (m_{J}+m_{D_{d}})\, A_{1}^{J{\to}D_{d}}
   \nonumber \\ & & \hspace{-15mm}
   + ({\epsilon}_{\rho}^{\ast}{\cdot}p_{J})\,
  ({\epsilon}_{J}{\cdot}p_{\rho})\,
   \frac{2\, A_{2}^{J{\to}D_{d}}}{ m_{J}+m_{D_{d}} }
  -i\,{\epsilon}_{{\mu}{\nu}{\alpha}{\beta}}\,
  {\epsilon}_{\rho}^{{\ast}{\mu}}\,{\epsilon}_{J}^{\nu}\,
  p_{\rho}^{\alpha}\,p_{J}^{\beta}\,
  \frac{2\, V^{J{\to}D_{d}}}{ m_{J}+m_{D_{d}} }
   \Big\}
   \label{amp-cd-rho}, \\
  {\cal A}(J/{\psi}{\to}D_{d}^{-}K^{{\ast}+})
  &=&
   -i\,\frac{G_{F}}{\sqrt{2}}\, f_{K^{\ast}}\, m_{K^{\ast}}\,
   V_{cd}^{\ast}\, V_{us}\, a_{1}\, \Big\{
   ({\epsilon}_{K^{\ast}}^{\ast}{\cdot}{\epsilon}_{J})\,
   (m_{J}+m_{D_{d}})\, A_{1}^{J{\to}D_{d}}
   \nonumber \\ & & \hspace{-15mm}
   + ({\epsilon}_{K^{\ast}}^{\ast}{\cdot}p_{J})\,
  ({\epsilon}_{J}{\cdot}p_{K^{\ast}})\,
   \frac{2\, A_{2}^{J{\to}D_{d}}}{ m_{J}+m_{D_{d}} }
  -i\,{\epsilon}_{{\mu}{\nu}{\alpha}{\beta}}\,
  {\epsilon}_{K^{\ast}}^{{\ast}{\mu}}\,{\epsilon}_{J}^{\nu}\,
  p_{K^{\ast}}^{\alpha}\,p_{J}^{\beta}\,
  \frac{2\, V^{J{\to}D_{d}}}{ m_{J}+m_{D_{d}} }
   \Big\}
   \label{amp-cd-kv}, \\
  {\cal A}(J/{\psi}{\to}\overline{D}_{u}^{0}{\rho}^{0})
  &=&
   +i\,\frac{G_{F}}{2}\, f_{\rho}\, m_{\rho}\,
   V_{cd}^{\ast}\, V_{ud}\, a_{2}\, \Big\{
   ({\epsilon}_{\rho}^{\ast}{\cdot}{\epsilon}_{J})\,
   (m_{J}+m_{D_{u}})\, A_{1}^{J{\to}D_{u}}
   \nonumber \\ & & \hspace{-15mm}
   + ({\epsilon}_{\rho}^{\ast}{\cdot}p_{J})\,
  ({\epsilon}_{J}{\cdot}p_{\rho})\,
   \frac{2\, A_{2}^{J{\to}D_{u}}}{ m_{J}+m_{D_{u}} }
  -i\,{\epsilon}_{{\mu}{\nu}{\alpha}{\beta}}\,
  {\epsilon}_{\rho}^{{\ast}{\mu}}\,{\epsilon}_{J}^{\nu}\,
  p_{\rho}^{\alpha}\,p_{J}^{\beta}\,
   \frac{2\, V^{J{\to}D_{u}}}{ m_{J}+m_{D_{u}} }
   \Big\}
   \label{amp-cu-rho}, \\
  {\cal A}(J/{\psi}{\to}\overline{D}_{u}^{0}{\omega})
  &=&
   -i\,\frac{G_{F}}{2}\, f_{\omega}\, m_{\omega}\,
   V_{cd}^{\ast}\, V_{ud}\, a_{2}\, \Big\{
   ({\epsilon}_{\omega}^{\ast}{\cdot}{\epsilon}_{J})\,
   (m_{J}+m_{D_{u}})\, A_{1}^{J{\to}D_{u}}
   \nonumber \\ & & \hspace{-15mm}
   + ({\epsilon}_{\omega}^{\ast}{\cdot}p_{J})\,
  ({\epsilon}_{J}{\cdot}p_{\omega})\,
   \frac{2\, A_{2}^{J{\to}D_{u}}}{ m_{J}+m_{D_{u}} }
  -i\,{\epsilon}_{{\mu}{\nu}{\alpha}{\beta}}\,
  {\epsilon}_{\omega}^{{\ast}{\mu}}\,{\epsilon}_{J}^{\nu}\,
  p_{\omega}^{\alpha}\,p_{J}^{\beta}\,
   \frac{2\, V^{J{\to}D_{u}}}{ m_{J}+m_{D_{u}} }
   \Big\}
   \label{amp-cu-omega}, \\
  {\cal A}(J/{\psi}{\to}\overline{D}_{u}^{0}{\phi})
  &=&
   -i\,\frac{G_{F}}{\sqrt{2}}\, f_{\phi}\, m_{\phi}\,
   V_{cs}^{\ast}\, V_{us}\, a_{2}\, \Big\{
   ({\epsilon}_{\phi}^{\ast}{\cdot}{\epsilon}_{J})\,
   (m_{J}+m_{D_{u}})\, A_{1}^{J{\to}D_{u}}
   \nonumber \\ & & \hspace{-15mm}
   + ({\epsilon}_{\phi}^{\ast}{\cdot}p_{J})\,
  ({\epsilon}_{J}{\cdot}p_{\phi})\,
   \frac{2\, A_{2}^{J{\to}D_{u}}}{ m_{J}+m_{D_{u}} }
  -i\,{\epsilon}_{{\mu}{\nu}{\alpha}{\beta}}\,
  {\epsilon}_{\phi}^{{\ast}{\mu}}\,{\epsilon}_{J}^{\nu}\,
  p_{\phi}^{\alpha}\,p_{J}^{\beta}\,
   \frac{2\, V^{J{\to}D_{u}}}{ m_{J}+m_{D_{u}} }
   \Big\}
   \label{amp-cu-phi}, \\
  {\cal A}(J/{\psi}{\to}\overline{D}_{u}^{0}K^{{\ast}0})
  &=&
    -i\,\frac{G_{F}}{\sqrt{2}}\, f_{K^{\ast}}\, m_{K^{\ast}}\,
   V_{cd}^{\ast}\, V_{us}\, a_{2}\, \Big\{
   ({\epsilon}_{K^{\ast}}^{\ast}{\cdot}{\epsilon}_{J})\,
   (m_{J}+m_{D_{u}})\, A_{1}^{J{\to}D_{u}}
   \nonumber \\ & & \hspace{-15mm}
   + ({\epsilon}_{K^{\ast}}^{\ast}{\cdot}p_{J})\,
  ({\epsilon}_{J}{\cdot}p_{K^{\ast}})\,
   \frac{2\, A_{2}^{J{\to}D_{u}}}{ m_{J}+m_{D_{u}} }
  -i\,{\epsilon}_{{\mu}{\nu}{\alpha}{\beta}}\,
  {\epsilon}_{K^{\ast}}^{{\ast}{\mu}}\,{\epsilon}_{J}^{\nu}\,
  p_{K^{\ast}}^{\alpha}\,p_{J}^{\beta}\,
   \frac{2\, V^{J{\to}D_{u}}}{ m_{J}+m_{D_{u}} }
   \Big\}
   \label{amp-cu-kvz}, \\
  {\cal A}(J/{\psi}{\to}\overline{D}_{u}^{0}\overline{K}^{{\ast}0})
  &=&
  -i\,\frac{G_{F}}{\sqrt{2}}\, f_{K^{\ast}}\,
  m_{K^{\ast}}\, V_{cs}^{\ast}\, V_{ud}\, a_{2}\, \Big\{
   ({\epsilon}_{K^{\ast}}^{\ast}{\cdot}{\epsilon}_{J})\,
   (m_{J}+m_{D_{u}})\, A_{1}^{J{\to}D_{u}}
   \nonumber \\ & & \hspace{-15mm}
   + ({\epsilon}_{K^{\ast}}^{\ast}{\cdot}p_{J})\,
  ({\epsilon}_{J}{\cdot}p_{K^{\ast}})\,
   \frac{2\, A_{2}^{J{\to}D_{u}}}{ m_{J}+m_{D_{u}} }
  -i\,{\epsilon}_{{\mu}{\nu}{\alpha}{\beta}}\,
  {\epsilon}_{K^{\ast}}^{{\ast}{\mu}}\,{\epsilon}_{J}^{\nu}\,
  p_{K^{\ast}}^{\alpha}\,p_{J}^{\beta}\,
   \frac{2\, V^{J{\to}D_{u}}}{ m_{J}+m_{D_{u}} }
   \Big\}
   \label{amp-cu-kvzb}.
  \end{eqnarray}
  \end{appendix}


   \begin{table}[h]
   \caption{The numerical values of the Wilson coefficients $C_{1,2}$
    and effective coefficients $a_{1,2}$ for the $J/{\psi}$ ${\to}$
    $D{\pi}$ decay at different scales, where $m_{c}$ $=$
    1.275 GeV.}
   \label{tab01}
   \begin{ruledtabular}
   \begin{tabular}{c|cc|cc|cc|cccc}
 & \multicolumn{2}{c|}{LO} & \multicolumn{2}{c|}{NLO}
 & \multicolumn{2}{c|}{NF} & \multicolumn{4}{c}{QCDF} \\ \cline{2-11}
 ${\mu}$ & $C_{1}$ & $C_{2}$ & $C_{1}$ & $C_{2}$
         & $a_{1}$ & $a_{2}$ & Re($a_{1}$) & Im($a_{1}$)
         & Re($a_{2}$) & Im($a_{2}$) \\ \hline
 $0.8\,m_{c}$ & $ 1.335$ & $-0.589$ & $ 1.275$ & $-0.504$
              & $ 1.107$ & $-0.079$
              & $ 1.271$ & $ 0.097$ & $-0.453$ & $-0.219$ \\
 $m_{c}$      & $ 1.276$ & $-0.505$ & $ 1.222$ & $-0.425$
              & $ 1.080$ & $-0.018$
              & $ 1.217$ & $ 0.069$ & $-0.363$ & $-0.173$ \\
 $1.2\,m_{c}$ & $ 1.239$ & $-0.450$ & $ 1.190$ & $-0.374$
              & $ 1.065$ & $ 0.022$
              & $ 1.185$ & $ 0.054$ & $-0.308$ & $-0.149$
   \end{tabular}
   \end{ruledtabular}
   \end{table}

   \begin{table}[h]
   \caption{The values of the Wolfenstein parameters,
   decay constants and Gegenbauer moments.}
   \label{tab02}
   \begin{ruledtabular}
  \begin{tabular}{ll}
  \multicolumn{2}{c}{Wolfenstein parameters} \\ \hline
    ${\lambda}$  $=$ $0.22537{\pm}0.00061$     \cite{pdg}
  & $A$          $=$ $0.814^{+0.023}_{-0.024}$ \cite{pdg} \\
    $\bar{\rho}$ $=$ $0.117{\pm}0.021$         \cite{pdg}
  & $\bar{\eta}$ $=$ $0.353{\pm}0.013$         \cite{pdg} \\ \hline
  \multicolumn{2}{c}{decay constants} \\ \hline
    $f_{\pi}$ $=$ $130.41{\pm}0.20$ MeV \cite{pdg}
  & $f_{K}  $ $=$ $156.2{\pm}0.7$ MeV \cite{pdg} \\
    $f_{{\eta}_{q}}$ $=$ $(1.07{\pm}0.02)f_{\pi}$ \cite{eta}
  & $f_{{\eta}_{s}}$ $=$ $(1.34{\pm}0.06)f_{\pi}$ \cite{eta} \\
    $f_{\rho}$   $=$ $216{\pm}3$ MeV \cite{ball3}
  & $f_{\omega}$ $=$ $187{\pm}5$ MeV \cite{ball3} \\
    $f_{\phi}$   $=$ $215{\pm}5$ MeV \cite{ball3}
  & $f_{K^{\ast}}$ $=$ $220{\pm}5$ MeV \cite{ball3} \\ \hline
  \multicolumn{2}{c}{Gegenbauer moments at the scale ${\mu}$ $=$ 1 GeV} \\ \hline
    $a_{1}^{\pi}$ $=$ $a_{1}^{{\eta}_{q}}$ $=$
    $a_{1}^{{\eta}_{s}}$ $=$ $0$ \cite{ball2}
  & $a_{2}^{\pi}$ $=$ $a_{2}^{{\eta}_{q}}$ $=$
    $a_{2}^{{\eta}_{s}}$ $=$ $0.25{\pm}0.15$ \cite{ball2} \\
    $a_{1}^{\bar{K}}$ $=$ $-a_{1}^{K}$ $=$ $0.06{\pm}0.03$ \cite{ball2}
  & $a_{2}^{K}$ $=$ $a_{2}^{\bar{K}}$ $=$ $0.25{\pm}0.15$ \cite{ball2} \\
    $a_{1}^{\rho}$ $=$ $a_{1}^{\omega}$ $=$ $a_{1}^{\phi}$ $=$ $0$ \cite{ball3}
  & $a_{2}^{\rho}$ $=$ $a_{2}^{\omega}$ $=$ $0.15{\pm}0.07$ \cite{ball3} \\
    $a_{1}^{\bar{K}^{\ast}}$ $=$ $-a_{1}^{K^{\ast}}$ $=$ $0.03{\pm}0.02$ \cite{ball3}
  & $a_{2}^{K^{\ast}}$ $=$ $a_{2}^{\bar{K}^{\ast}}$ $=$ $0.11{\pm}0.09$ \cite{ball3} \\
  & $a_{2}^{\phi}$ $=$ $0.18{\pm}0.08$ \cite{ball3}
  \end{tabular}
  \end{ruledtabular}
  \end{table}

   {\renewcommand{\baselinestretch}{1.2}
   \begin{sidewaystable}[h]
   \caption{The branching ratios for $J/{\psi}$ ${\to}$ $DP$, $DV$ decay,
    The numbers in columns of Refs.\cite{epjc55,adv2013,ijma14} are
    calculated with coefficients $a_{1}$ = 1.26 and $a_{2}$ $=$ $-0.51$.
    The results of Ref.\cite{epjc55} are based on QCD sum rules.
    The numbers in columns of ``A'', ``B'', ``C'' and ``D'' are based
    on BSW model with flavor dependent ${\omega}$, QCD inspired
    ${\omega}$ $=$ ${\alpha}_{s}$ ${\times}$ $m$, universal
    ${\omega}$ $=$ 0.4 GeV and 0.5GeV, respectively.
    The uncertainties of the ``QCDF'' column come from the CKM
    parameters, the renormalization scale ${\mu}$ $=$ $(1{\pm}0.2)m_{c}$,
    decay constants
    and Gegenbauer moments, form factors, respectively.}
   \label{tab03}
   \begin{ruledtabular}
   \begin{tabular}{l|c|c|c|c|c|c|c}
   Decay & & Ref.\cite{epjc55}
           & \multicolumn{3}{c|}{ Ref.\cite{adv2013} }
           & Ref.\cite{ijma14}
           & This work \\ \cline{4-8}
   modes  & Case & & A & B & C & D & QCDF \\ \hline
 $D_{s}^{-}{\pi}^{+}$
  & 1 a
  & $2.0 {\times}10^{-10}$
  & $7.41{\times}10^{-10}$
  & $7.13{\times}10^{-10}$
  & $3.32{\times}10^{-10}$
  & $8.74{\times}10^{-10}$
  & $(4.10^{+0.00+0.39+0.02+1.63}_{-0.00-0.22-0.02-1.35}){\times}10^{-10}$
 \\
 $D_{s}^{-}K^{+}$
  & 1 b
  & $1.6{\times}10^{-11}$
  & $5.3{\times}10^{-11}$
  & $5.2{\times}10^{-11}$
  & $2.4{\times}10^{-11}$
  & $5.5{\times}10^{-11}$
  & $(2.32^{+0.01+0.22+0.03+0.92}_{-0.01-0.12-0.03-0.77}){\times}10^{-11}$
 \\
 $D_{d}^{-}{\pi}^{+}$
  & 1 b
  & $0.8{\times}10^{-11}$
  & $2.9{\times}10^{-11}$
  & $2.8{\times}10^{-11}$
  & $1.5{\times}10^{-11}$
  & $5.5{\times}10^{-11}$
  & $(2.21^{+0.01+0.21+0.01+0.97}_{-0.01-0.12-0.01-0.79}){\times}10^{-11}$
 \\
 $D_{d}^{-}K^{+}$
  & 1 c
  & ------
  & $2.3{\times}10^{-12}$
  & $2.2{\times}10^{-12}$
  & $1.2{\times}10^{-12}$
  & ------
  & $(1.31^{+0.01+0.13+0.02+0.58}_{-0.01-0.07-0.02-0.47}){\times}10^{-12}$
 \\
 $\overline{D}_{u}^{0}{\pi}^{0}$
  & 2 b
  & ------
  & $2.4{\times}10^{-12}$
  & $2.3{\times}10^{-12}$
  & $1.2{\times}10^{-12}$
  & $5.5{\times}10^{-12}$
  & $(1.21^{+0.01+0.69+0.02+0.53}_{-0.01-0.34-0.02-0.44}){\times}10^{-12}$
 \\
 $\overline{D}_{u}^{0}K^{0}$
  & 2 c
  & ------
  & $4.0{\times}10^{-13}$
  & $4.0{\times}10^{-13}$
  & $2.0{\times}10^{-13}$
  & ------
  & $(1.44^{+0.02+0.81+0.04+0.63}_{-0.02-0.40-0.04-0.52}){\times}10^{-13}$
 \\
 $\overline{D}_{u}^{0}\overline{K}^{0}$
  & 2 a
  & $3.6{\times}10^{-11}$
  & $1.39{\times}10^{-10}$
  & $1.34{\times}10^{-10}$
  & $7.2{\times}10^{-11}$
  & $2.8{\times}10^{-10}$
  & $(4.98^{+0.00+2.81+0.12+2.19}_{-0.00-1.38-0.11-1.79}){\times}10^{-11}$
 \\
 $\overline{D}_{u}^{0}{\eta}$
  & 2 b
  & ------
  & $7.0{\times}10^{-12}$
  & $6.7{\times}10^{-12}$
  & $3.6{\times}10^{-12}$
  & $1.6{\times}10^{-12}$
  & $(3.56^{+0.02+2.01+0.24+1.57}_{-0.02-0.99-0.29-1.28}){\times}10^{-12}$
 \\
 $\overline{D}_{u}^{0}{\eta}^{\prime}$
  &
  & ------
  & $4.0{\times}10^{-13}$
  & $4.0{\times}10^{-13}$
  & $2.0{\times}10^{-13}$
  & $3.0{\times}10^{-13}$
  & $(2.02^{+0.01+1.14+0.23+0.89}_{-0.01-0.56-0.41-0.73}){\times}10^{-13}$
 \\ \hline
 $D_{s}^{-}{\rho}^{+}$
  & 1 a
  & $1.26{\times}10^{-9}$
  & $5.11{\times}10^{-9}$
  & $5.32{\times}10^{-9}$
  & $1.77{\times}10^{-9}$
  & $3.63{\times}10^{-9}$
  & $(2.21^{+0.00+0.21+0.06+0.78}_{-0.00-0.12-0.06-0.66}){\times}10^{ -9}$
 \\
 $D_{s}^{-}K^{{\ast}+}$
  & 1 b
  & $0.82{\times}10^{-10}$
  & $2.82{\times}10^{-10}$
  & $2.96{\times}10^{-10}$
  & $0.97{\times}10^{-10}$
  & $2.12{\times}10^{-10}$
  & $(1.22^{+0.01+0.11+0.06+0.42}_{-0.01-0.06-0.06-0.36}){\times}10^{-10}$
 \\
 $D_{d}^{-}{\rho}^{+}$
  & 1 b
  & $0.42{\times}10^{-10}$
  & $2.16{\times}10^{-10}$
  & $2.28{\times}10^{-10}$
  & $0.72{\times}10^{-10}$
  & $2.20{\times}10^{-10}$
  & $(1.09^{+0.01+0.10+0.03+0.45}_{-0.01-0.06-0.03-0.37}){\times}10^{-10}$
 \\
 $D_{d}^{-}K^{{\ast}+}$
  & 1 c
  & ------
  & $1.3{\times}10^{-11}$
  & $1.3{\times}10^{-11}$
  & $4.2{\times}10^{-12}$
  & ------
  & $(6.14^{+0.07+0.58+0.30+2.51}_{-0.07-0.32-0.29-2.08}){\times}10^{-12}$
 \\
 $\overline{D}_{u}^{0}{\rho}^{0}$
  & 2 b
  & ------
  & $1.8{\times}10^{-11}$
  & $1.9{\times}10^{-11}$
  & $6.0{\times}10^{-12}$
  & $2.2{\times}10^{-11}$
  & $(5.93^{+0.03+3.35+0.20+2.45}_{-0.03-1.64-0.20-2.03}){\times}10^{-12}$
 \\
 $\overline{D}_{u}^{0}{\omega}$
  & 2 b
  & ------
  & $1.6{\times}10^{-11}$
  & $1.7{\times}10^{-11}$
  & $5.0{\times}10^{-12}$
  & $1.8{\times}10^{-11}$
  & $(4.45^{+0.02+2.51+0.27+1.84}_{-0.02-1.23-0.26-1.52}){\times}10^{-12}$
 \\
  $\overline{D}_{u}^{0}{\phi}$
  & 2 b
  & ------
  & $4.2{\times}10^{-11}$
  & $4.4{\times}10^{-11}$
  & $1.4{\times}10^{-11}$
  & $6.5{\times}10^{-11}$
  & $(1.11^{+0.01+0.63+0.06+0.45}_{-0.01-0.31-0.06-0.37}){\times}10^{-11}$
 \\
 $\overline{D}_{u}^{0}K^{{\ast}0}$
  & 2 c
  & ------
  & $2.1{\times}10^{-12}$
  & $2.2{\times}10^{-12}$
  & $7.0{\times}10^{-13}$
  & ------
  & $(6.69^{+0.07+3.78+0.38+2.74}_{-0.07-1.85-0.36-2.27}){\times}10^{-13}$
 \\
 $\overline{D}_{u}^{0}\overline{K}^{{\ast}0}$
  & 2 a
  & $1.54{\times}10^{-10}$
  & $7.61{\times}10^{-10}$
  & $8.12{\times}10^{-10}$
  & $2.51{\times}10^{-10}$
  & $1.03{\times}10^{-9}$
  & $(2.32^{+0.00+1.31+0.13+0.95}_{-0.00-0.64-0.12-0.79}){\times}10^{-10}$
  \end{tabular}
  \end{ruledtabular}
  \end{sidewaystable}
  }

  \end{document}